\documentclass[12pt]{iopart1}

\usepackage{amsmath}
\usepackage{amssymb}
\usepackage{color}
\usepackage{graphicx}   
\usepackage{hyperref}

\newcommand\al{\alpha}

\newcommand\de{\delta}
\newcommand\ep{\epsilon}
\newcommand\ze{\zeta}
\newcommand\et{\eta}
\renewcommand\th{\theta}

\newcommand\ka{\kappa}
\newcommand\la{\lambda}
\newcommand\rh{\rho}
\newcommand\si{\sigma}

\newcommand\ph{\phi}
\newcommand\vp{\varphi}

\newcommand\ps{\psi}

\newcommand\pa{\partial}

\newcommand\<{\langle}
\renewcommand\>{\rangle}

\newcommand\ie{{\em i.e.}}

\newcommand\beq{\begin{equation}}
\newcommand\eeq{\end{equation}}
\newcommand\bea{\begin{eqnarray}}
\newcommand\eea{\end{eqnarray}}
\newcommand\bal{\begin{align}}
\newcommand\eal{\end{align}}
\newcommand\bpm{\begin{pmatrix}}
\newcommand\epm{\end{pmatrix}}

\newcommand\X{\times}
\newcommand\fr{\frac}

\newcommand\rms[1]{_{\mathrm{#1}}}

\newcommand\half{\tfrac{1}{2}}

\newcommand\cd{\cdot}

\newcommand\bone{\boldsymbol{1}}

\newcommand\cL{\mathcal{L}}
\newcommand\cM{\mathcal{M}}
\newcommand\cV{\mathcal{V}}
\newcommand\CC{\mathbb{C}}
\newcommand\RR{\mathbb{R}}
\newcommand\ZZ{\mathbb{Z}}

\begin{document}
\title{Monopoles on strings}

\author{T W B Kibble$^1$, Tanmay Vachaspati$^2$}

\address{$^1$ Blackett Laboratory, Imperial College, London SW7~2AZ, UK}
\address{$^2$ Department of Physics, Arizona State University, Tempe, AZ 85287}
\ead{\mailto{kibble@ic.ac.uk},\mailto{tvachasp@asu.edu}}

\begin{abstract}
In cosmological scenarios based on grand unification, string theory or braneworlds, many kinds of topological or non-topological defects, including monopoles and cosmic strings, are predicted to be formed in the early universe.  Here we review specifically the physics of composite objects involving monopoles tied to strings.  There is a wide variety of these, including for example ``dumbbells'' and ``necklaces'', depending on how many strings attach to each monopole and on the extent to which the various fluxes are confined to the strings.  We also briefly survey the prospects for observing such structures, the existing observational limits, and potential evidence for a cosmological role.
\end{abstract}
\submitto{\jpg}
\maketitle

\section{Introduction}

Symmetry-breaking phase transitions at which topological defects form are ubiquitous in low-temperature condensed-matter systems.  In fundamental particle physics too, there are good reasons for thinking that similar phenomena occur at vastly higher energy scales.  If so, the defects formed may have significant effects in the early universe.

Electroweak unification is now well-established.  It is natural to suppose that further unification may occur at even higher energies, in a Grand Unified Theory (GUT).  Some evidence in support of this idea comes from the running of the three independent coupling constants in the $SU(3)\X SU(2)\X U(1)$ standard model.  In the basic standard model, they do not quite seem to meet.  But in its supersymmetric extension, the parameters can be chosen so that they do all come together at an energy of around $10^{15}$ GeV \cite{Amaldi:1991zx,Haber:1997if}, suggesting the possibility of a high-temperature phase with a larger symmetry represented by a simple group such as $SO(10)$.  In such a model, the universe would be expected to go through one or more phase transitions with decreasing symmetry as it cooled after the Big Bang.  Even without supersymmetry, models with multiple phase transitions may be viable \cite{Mambrini:2015vna}.

Depending on the pattern of symmetry breaking, such transitions could create topological defects of various types.  These include point defects (monopoles), linear defects (cosmic strings, analogous to vortices in condensed matter) and domain walls.  It has long been known that because the standard model group contains a $U(1)$ factor, monopoles are generic in GUT models that start with a simple gauge group.  Avoiding the resulting over-abundance of monopoles was one of the prime original motivations for the introduction of the theory of inflation; if the monopoles are generated \emph{before} the inflationary era, they will be diluted to insignificance by the rapid expansion.  Nevertheless, inflation is compatible with the existence of defects, which can be formed during the reheating phase that terminates that era \cite{Sarangi:2002yt}.  Moreover, essentially all realistic GUTs predict the existence of cosmic strings, though not always stable ones \cite{Jeannerot:2003qv}.  

Cosmological scenarios derived from fundamental string theory or M-theory, such as braneworld models, also frequently predict the appearance of defects of similar types \cite{Jones:2002cv,Majumdar:2002hy,Sarangi:2002yt,Becker:2005pv}.  These strings can have somewhat different properties.  In particular, the probability of exchanging partners when strings intersect, can be much less than one \cite{Jackson:2004zg}, in contrast to the situation for cosmic strings in gauge theories \cite{Hashimoto:2005hi}.  Moreover, there can be strings of different tension, fundamental strings (F-strings) as well as Dirichlet D1 branes (D-strings) and ($p,q$)-strings, composites of $p$ F-strings and $q$ D-strings \cite{Polchinski:2004ia,Dvali:2003zj}.  There may be junctions where three strings meet.  The evolution of a network of such strings is a more complicated problem, but the final result may not be so very different \cite{Avgoustidis:2005nv}.  Analytic and numerical studies have shown that a network of ordinary cosmic strings generally evolves, at least on large scales, to a scaling regime in which the strings form a roughly constant fraction of the energy density of the Universe.  Though the analysis is less clear cut, this appears also to be true for a multi-tension network.  In that case the lightest strings come to dominate \cite{Tye:2005fn,Hindmarsh:2006qn}.  Here we shall not discuss these added complications in any detail. 
Another important topic that lies outside the scope of this review is the relevance of monopole and string networks to QCD confinement.  See for example  \cite{Kneipp:2007fg}.

There has been extensive discussion in the literature of the characteristics and effects of these various defects, and of ways in which they might be detected.  But in addition to the simple defects there may also be composite defects such as domain walls bounded by strings and strings connected to monopoles, and there has been less discussion of the effects of these more exotic structures.  In this paper, we shall concentrate on the composites of strings and monopoles.  As we aim to make clear, even within this category there are many different types of structures.

In Sec.~\ref{simpledefects}, we briefly review the topological requirements for the different types of defects to form at a phase transition, where the symmetry is broken from a group $G$ to a subgroup $H$.  These are governed by the topology, in particular the homotopy groups, of the manifold $\cM$ of degenerate vacuum states, which may be identified with the quotient space $G/H$.  Then in Sec.~\ref{compositedefects}, we discuss the case where the system undergoes two successive phase transitions, with the symmetry first broken from $G$ to $H$ and then to a smaller subgroup $K$.  Such scenarios often lead to the formation of composite defects.  In the remainder of the section, we discuss a number of different models that illustrate the wide range of possible defect structures.  The interactions between the various defects that can form can be quite complex \cite{Sakai:2005sp,Tong:2005un}.  Here we concentrate only on a few examples.

Sec.~\ref{stdmodel} is devoted to a discussion of the strings and monopoles that appear in the standard electroweak model.  Electroweak monopoles and strings are not strictly speaking topological defects. Electroweak monopoles are confined while electroweak strings are known to be unstable, but configurations of these electroweak defects still can play an important role in cosmology, especially perhaps in connection with baryon number violation and cosmological magnetic field generation.

Possible means of observing composite defects of various kinds are discussed in Sec.~\ref{observations}, where we also discuss the observational constraints arising from existing observations.  The conclusions are briefly summarized in Sec.~\ref{conclusions}.

\section{Simple defects \label{simpledefects}}

\subsection{Topological conditions for defects}

We first recall the conditions for the appearance of topological defects of various types at a symmetry-breaking phase transition (see for example \cite{Kibble:2002ex}).  When the system is cooled through the transition temperature,  there is some order parameter field multiplet $\ph$ that acquires a vacuum expectation value, say $\<0|\ph|0\>=\ph_0$, lying somewhere on a manifold $\cM$ of minima of the potential $\cV(\ph)$.  If the symmetry group is $G$, then any operation $g\in G$ will transform this vacuum state into another degenerate vacuum state, with expectation value $g\ph_0$.  If $H=\{h\in G|h\ph_0=\ph_0\}\subset G$ is the subgroup leaving $\ph_0$ invariant, then the vacuum manifold may be identified with the quotient $\cM=G/H$, the set of left cosets $\{gH\}$ of $H$ in $G$.  The types of defects that may be formed are governed by the topology of $\cM$.

\emph{Cosmic strings} can form if $\cM$ is not simply connected, \ie \ its fundamental group or first homotopy group $\pi_1(\cM)\ne 1$, where $1$ stands for the group comprising the identity alone.  That means there are closed loops in $\cM$ that cannot be continuously shrunk to a point.  The value of $\ph$ at points on a large loop in space surrounding a cosmic string will follow such a path.  In the simplest case, where $\ph$ is a complex scalar and $G$ comprises the phase rotations $\ph\to\ph e^{i\al}$, then $\cM$ is a circle, $|\ph|=\et$, \ie\ $\ph=\et e^{i\al}$ with arbitrary phase $\al$.  In this case, $\pi_1(\cM)=\ZZ$.  Thus the strings are labelled by an integer \emph{winding number} $n$; on a loop around a string of winding number $n$, the phase $\al$ changes by $2n\pi$ (see section \ref{strings}).

There is a simple general criterion for the existence of strings in a model where the symmetry group $G$ is connected and simply connected, \ie\ $\pi_0(G)=\pi_1(G)=1$.  Then a standard theorem tells us that $\pi_1(G/H)=\pi_0(H)$.  Here the zeroth homotopy group counts the number of disconnected pieces of $H$; $\pi_0(H)=H/H_0$, where $H_0$ is the connected component of $H$ containing the identity.   So strings exist if and only if $H$ is disconnected.  The theorem may still be applied even if $G$ is not simply connected, merely by replacing it by its simply connected universal covering group.  For example, we may replace $U(1)$ by the additive group of real numbers $G=\RR$, in which case $H=\ZZ$, the set of transformations with $\al=2n\pi$. 

Similarly, \emph{monopoles} exist if $\pi_2(\cM)\ne 1$, that is, if there are non-shrinkable two-dimensional surfaces in $\cM$.  Surrounding a monopole, the value of $\ph$ will lie on such a surface.  The simplest example here is when $G=SU(2)$, with $\ph$ in the three-dimensional adjoint representation.  Then $H=U(1)$, and $\cM$ is a 2-sphere, $|\ph|=\et$, so $\pi_2(\cM)=\ZZ$.  The monopoles are again labelled by an integer (see section \ref{tHPol}).  There is also a similar theorem, applicable when $\pi_0(G)=\pi_1(G)=1$, namely $\pi_2(G/H)=\pi_1(H)$.  So monopoles exist when $H$ is not simply connected.

For completeness, we mention two other topological objects.  \emph{Domain walls} occur when $\cM$ itself is disconnected, $\pi_0(\cM)\ne1$.  For example, we may take a real scalar field $\ph$ with a double-well potential and $\ZZ_2$ symmetry under $\ph\to-\ph$.  The domain wall separates regions where the vacuum expectation value lies in one well or the other.  Finally \emph{textures} occur if $\pi_3(\cM)\ne1$.  Here there is no actual defect, in the sense of a compact region of concentrated energy.  However, a non-trivial texture cannot be smoothly eliminated and converted to the vacuum state; it represents excess energy, albeit spread out rather than concentrated.  Unwinding of the texture occurs in a restricted region of spacetime.  Textures in the universe could have real physical consequences \cite{Turok:1989ai}.

In the remainder of this section, we discuss specific models that illustrate a variety of different types of strings and monopoles.

\subsection{Simple string models \label{strings}}

{\bf a.}  The simplest model that leads to cosmic strings is the $U(1)$-symmetric \emph{Abelian Higgs model}, comprising a complex scalar field $\ph$ interacting with a gauge field $A_\mu$, described by the Lagrangian 
 \beq \cL = D_\mu\ph^* D^\mu\ph - \tfrac{1}{4}F_{\mu\nu}F^{\mu\nu} - \cV, \eeq
where 
 \beq D_\mu\ph = \pa_\mu\ph+ieA_\mu\ph, \qquad F_{\mu\nu} = \pa_\mu A_\nu-\pa_\nu A_\mu, \eeq
and
 \beq \cV = \tfrac{1}{4}\la(\ph^*\ph - \eta^2)^2. \eeq
Here $\la$ and $\et$ are real positive constants and we set $c=\hbar=1$.  The potential $\cV$ has a maximum at $\ph=0$, so the $U(1)$ symmetry is broken in the vacuum.  There is a degenerate family of vacua labelled by the phase angle: $\<0|\ph|0\> = \et e^{i\al}$.  This is essentially scalar electrodynamics but with a symmetry-breaking potential.  The masses of the scalar and vector particles in the theory are $m\rms{s}=\sqrt{\la}\et$, $m\rms{v}=\sqrt{2}|e|\et$.

A static string with winding number $n$ along the $z$ axis is described in cylindrical polars $(\rh,\vp,z)$ by a field configuration
 \beq \ph = \et f(\rh) e^{ni\vp}, \qquad A_0=0,\qquad A_k=-\fr{n}{e\rh} h(\rh) \pa_k\vp, \label{NO}\eeq
where the dimensionless functions $f$ and $h$ satisfy the boundary conditions
 \beq f(0)=h(0)=0,\qquad f(\infty)=h(\infty)=1. \eeq
The magnetic field along the string carries a total magnetic flux $2n\pi/e$.  This is the \emph{Nielsen--Olesen string} solution \cite{Nielsen:1973cs}.

The solution with $n=1$ is always stable, but the stability of strings with $n>1$ depends on the value of the ratio $\beta=m\rms{s}^2/m\rms{v}^2=\la/2e^2$.  For Type-II strings, with $\beta>1$, close parallel strings repel, and any string with $n>1$ is unstable to break-up into $n=1$ strings.  Type-I strings, with $\beta<1$, are stable for all values of $n$, and can form three-string junctions where for example strings with winding numbers $m$ and $n$ meet to form an $(m+n)$ string.  For the critical case of $\beta=1$, there is no force between parallel strings.

For these strings, the tension is equal to the energy per unit length, $\mu$, and is given by $\mu=2\pi g(\beta)\et^2$, where $g$ is a slowly varying, monotonically increasing function with the value $g(1)=1$ for the critical coupling.

{\bf b.}  As a second example, we consider the symmetry group $G=SU(2)$ with \emph{two} scalar fields in the adjoint representation, $\ph=\ph^a\si^a$ and $\ps=\ps^a\si^a$, where the $\si^a$ are Pauli matrices.  If we take 
 \beq \cV = \tfrac{1}{4}\la({\vec\ph}^2-\et^2)^2+\tfrac{1}{4}\la({\vec\ps}^2-\et^2)^2
 +\tfrac{1}{4}\mu(\vec\ph\cd\vec\ps)^2, \label{V2vec} \eeq
where ${\vec\ph}^2=\ph^a\ph^a=\half\tr(\ph^2)$, then it is clear that in the vacuum we will have $|\vec\ph|=|\vec\ps|=\et$ and $\vec\ph\cd\vec\ps=0$.  This breaks the symmetry down to the centre of $SU(2)$, namely $H=\{1,-1\}\cong\ZZ_2$.  Since $H$ is discrete, there is no remaining massless gauge field.  But there are strings, because $\pi_1(\cM)=\pi_0(H)=\ZZ_2$.  For these ``$\ZZ_2$ strings'', stable strings with higher winding numbers may exist but are not stable for topological reasons.  We shall return to this point below.

In fact, it is easy to construct similar models of ``$\ZZ_n$ strings'', by taking the symmetry group to be $SU(n)$ with $n$ fields in the adjoint representation, and choosing a potential that constrains them all to be non-zero and mutually orthogonal (see Section \ref{mono-str}d).

{\bf c.}  Interesting possibilities occur when some of the symmetries are local and others global \cite{Achucarro:1999it}.  For example, suppose that $G=U(2) \cong SU(2)\X U(1)/\ZZ_2$, with a scalar field in the fundamental (spinor) representation.  (The $\ZZ_2$ factor is required because the centre of $SU(2)$, comprising the two elements $\{1,-1\}$ is also contained in $U(1)$.)  Moreover, suppose that only the Abelian factor $U(1)$ is gauged, so there is just one gauge boson, while $SU(2)$ is a \emph{global} symmetry group.

If we looked only at the local symmetry, we might expect the appearance of strings because of the breaking of $U(1)$.  However the vacuum manifold $\ps^\dag\ps=\et^2$ is $\cM=S^3$, and $\pi_1(\cM)=1$, so there are no \emph{topologically stable} strings.  It is easy to construct a string solution by embedding the Nielsen--Olesen string solution (\ref{NO}); we take
 \beq \ph = \et f(\rh) e^{ni\vp}\bpm 0\\ 1 \epm, \qquad A_0=0,\qquad 
 A_k=-\fr{n}{e\rh} h(\rh) \pa_k\vp. 
 \label{semilocalstringsoln}
 \eeq
Stability of this \emph{semi-local} solution is not guaranteed by any topological argument.  Using the other 
component of $\ph$, it can be smoothly deformed into a configuration lying entirely in $\cM$, so that the 
potential energy vanishes, but at the cost of increasing the gradient energy.  Despite the absence of a 
topological guarantee of stability, detailed analysis shows that it is indeed dynamically stable in the Type-I regime 
$\beta<1$, though not when $\beta>1$ \cite{Hindmarsh:1991jq,James:1992wb}.

\subsection{'t Hooft--Polyakov monopoles \label{tHPol}}

In this model, $G=SU(2)$, and $\ph$ belongs to the three-dimensional adjoint representation.  We can write $\ph=\ph^a\si^a$, $A_\mu=A_\mu^a\si^a$.  Here we take
 \beq D_\mu\ph = \pa_\mu\ph + \half ie[A_\mu,\ph],\qquad
 F_{\mu\nu} = \pa_\mu A_\nu-\pa_\nu A_\mu + \half ie[A_\mu,A_\nu], \eeq
or equivalently
 \beq D_\mu\ph^a = \pa_\mu\ph^a - e\ep^{abc}A^b_\mu\ph^c,\qquad
 F^a_{\mu\nu} = \pa_\mu A^a_\nu-\pa_\nu A^a_\mu - e\ep^{abc}A^b_\mu A^c_\nu. \eeq
With $\cV=\tfrac{1}{4}(\vec\ph^2-\et^2)^2$, we find that the vacuum manifold $\cM$ is a two-sphere, $\vec\ph^2=\et^2$, and $H=U(1)$.  In this case there is a scalar particle of mass $m\rms{s}=\sqrt{\la}\et$ and three vector particles, one massless (identified with the photon) and two with masses $m\rms{v}=\sqrt{2}|e|\et$.  This is essentially the Weinberg-Salam model with vanishing weak mixing angle, $\th\rms{w}=0$.

Here a static monopole at the origin is described by the solution \cite{'tHooft:1974qc,Polyakov:1974ek}
 \beq \ph^a = \et f(r)\fr{x^a}{r},\qquad A^a_0=0, \qquad A^a_k = -h(r)\fr{\ep^{akj}x^j}{er^2}, \eeq
where $r=\sqrt{x^kx^k}$ and the functions $f$ and $h$ obey the boundary conditions
 \beq f(0)=h(0)=0, \qquad f(\infty)=h(\infty)=1. \eeq
At large values of $r$, the gauge field is found to be
 \beq F^a_{0k}=0, \qquad F^a_{ij}=\fr{x^a}{r}\fr{\ep_{ijk}x^k}{er^3}. \eeq
This field is in the direction of the unbroken symmetry generator, corresponding to the electromagnetic field.  It represents a radial magnetic field
 \beq B^k \equiv -\frac{1}{2} \epsilon^{kij} {\hat \phi}^a F_{ij}^a = - \fr{x^k}{er^3}. \eeq
Hence the total outward magnetic flux, the magnetic charge of the monopole, is
 \beq q=-\fr{4\pi}{e}. \eeq

In any model, for the field $A^a_k$ to be single-valued, the magnetic charge for any monopole must always satisfy the condition
 \beq eq = 2n\pi \eeq
for some integer $n$.  Note that for this particular monopole solution, the charge is twice the minimal value.

It can be shown that the mass of the monopole obeys the Bogomol'nyi bound \cite{Bogomolnyi:1976},
 \beq m\rms{mon} \ge \fr{4\pi\et}{e}, \eeq
which is saturated in the Prasad-Sommerfeld limit of small scalar coupling, $\la/e^2\to 0$ \cite{Prasad:1975kr}.

\section{Composite defects \label{compositedefects}}

\subsection{Defects formed at multiple phase transitions}

There are many field-theory models that predict more than one phase transition in the early universe.    In such cases, composite defects may form \cite{Kibble:1982dd}.  

Suppose we start with a theory with symmetry group $G$, and that it goes through a phase transition where a field $\ph$ acquires a non-zero expectation value, breaking the symmetry to a subgroup $H\subset G$, and then subsequently a second phase transition, where another field $\ps$ gets a non-zero, but generally smaller, 
vacuum expectation value, breaking the symmetry further to $K\subset H$.  After the first breaking, we have a vacuum manifold $\cM = G/H$.  If this is topologically non-trivial, defects will form.  In the second transition another set of defects may form if the manifold $H/K$ has non-trivial homotopy groups.  However, the existence of stable defects in the final phase is actually controlled by the topology of $\cM' = G/K$.

The simplest example here is a $U(1)$ gauge model with two scalar fields, $\ph$ of charge $2e$, and $\ps$ of charge $e$.  We assume that the potential contains an interaction term of the form $-m(\ph^*\ps^2+\ps^{*2}\ph)$.  The absolute minimum of the potential occurs when $\ph$ and $\ps$ have fixed magnitudes, say $|\ph|=\et,|\ps|=\ze$, and the phase of $\ps^2$ is the same as that of $\ph$.  After the first stage of symmetry breaking, when $\<\ph\>$ becomes non-zero, the symmetry is reduced from $U(1)$ to $H=\ZZ_2$, comprising the transformation $\ps\to-\ps$.   Here $\cM$ is a circle $S^1$, and $\pi_1(\cM)=\ZZ$.  Therefore strings are formed, labelled by an integer winding number $n$, with the phase of $\ph$ changing by $2n\pi$ around the string.

Now when the second transition occurs, the remaining $\ZZ_2$ symmetry is broken, because $\ps$ has to choose between the two degenerate vacuum values.  Breaking this discrete symmetry would be expected to create domain walls, separate regions where opposite choices are made.  But considering the overall symmetry breaking, from $U(1)$ to 1, no discrete symmetry breaking is involved, and there are no truly stable domain walls.  In fact, $\cM'$ is also a circle, but each point of $\cM$ corresponds to two diametrically opposite points of $\cM'$.

Consider a string along the $z$ axis, where outside the core, $\ph=\et e^{in\vp}$.  To minimize the potential we must then have $\ps=\pm\ze e^{in\vp/2}$.  But note that for $n=1$ or any odd number, that would not lead to a continuous solution.  The strings with even winding number survive, but around one with odd winding number there must be a point where $\ps$ changes sign over a short distance.  In other words, the string becomes attached as the boundary of a domain wall.

Unlike fully stable domain walls, these are potentially unstable to the formation of holes surrounded by new loops of string, though such a decay has to overcome an energy barrier.  The hole has to attain a minimum size before its creation becomes energetically favorable.

\subsection{Monopoles joined by strings \label{mono-str}}

We now discuss several examples where the first stage of the symmetry breaking leads to the formation of 't Hooft--Polyakov monopoles, followed by a second stage where strings form.

{\bf a.}  One simple example is provided by the $SU(2)$ model above with two adjoint fields $\ph$ and $\ps$, but where the constants $\la$ and $\et$ in the first two terms of (\ref{V2vec}) are different, say $\la,\la'$ and $\et,\et'$, with $\sqrt{\la'}\et'\ll\sqrt{\la}\et$.  Then in the first stage of symmetry breaking, when $\ph$ becomes non-zero, the symmetry will break to $H=U(1)$, while after the second stage it will break further to $K=\ZZ_2$.  At the first stage, monopoles will form, because $\pi_2(G/H)=\ZZ$.  In the second breaking, since $\pi_1(H/K)=\ZZ$, we expect strings, classified as usual by an integer winding number.  Overall, however, since $\pi_1(G/K)=\ZZ_2$, the only topologically stable strings are $\ZZ_2$ strings.

Moreover, there are no truly stable monopoles.  It is easy to see what happens.  Around an $n=1$ monopole the field $\vec\ph$ may be chosen to point radially outwards.  When $\vec\ps$ becomes nonzero it needs to be orthogonal to $\vec\ph$, so around a sphere it should lie in a tangential direction.  But it is not possible to choose such a direction everywhere.  There have to be points where it vanishes.  For example, we could take it everywhere in the azimuthal $\vp$ direction, but to maintain continuity it must then vanish at the north and south poles.  In fact, there have to be two strings attached to the monopole.  The monopoles are like beads on the string.  The configuration is often called a \emph{necklace}.  Similar structures can appear very naturally in string-theory models \cite{Leblond:2007tf}.

It is useful to consider the fields around a string.  If the first field $\vec\ph$ is taken to be along the string, then $\vec\ps$ must wind around it, either clockwise or anticlockwise.  Thus the string has a direction; a string is not identical to an anti-string, in spite of the fact that they are topologically equivalent.  A string can be converted to an antistring, but it takes energy to do so.  In fact, what it takes is the creation of a pair of monopoles.

Similarly, strings with higher winding numbers (in the Type-I case $\beta<1$) may exist, but are not truly stable; an $n=2$ string can terminate on a monopole.

{\bf b.}  Now let us consider another model, this time with symmetry group $G=U(2)$, as in the example of the semi-local string, but here with all symmetries gauged.  A scalar field $\ph$ in the adjoint representation breaks the symmetry, here to $H=U(1)\X U(1)$.  The manifold of degenerate vacua is $\cM = S^2$, and monopoles can form.  Now suppose there is another scalar field $\ps$ in the fundamental (spinor) representation, and that there are extra terms in the potential:
 \beq \cV = \tfrac{1}{4}\la(\ph^a\ph^a-\et^2)^2 + \half\la'(\ps^\dag\ps-\et'^2)^2 
 + g\ps^\dag\si^a\ps\ph^a, 
 \label{phipsimodel}
 \eeq
where again $\sqrt{\la'}\et'\ll\sqrt{\la}\et$.  As the system cools further, it will go through a second transition at which $\<\ps\>$ becomes non-zero.  If, for example, $\<\ph^a\>=\et\de^a_3$, then clearly, to minimize the potential, $\<\ps\>$ should be proportional to the eigenvector $0\choose1$ of $\si^3$.  This breaks the symmetry down to $K=U(1)$, generated by $\half(1+\si^3)={{1\,0}\choose{0\,0}}$.

This model is very different from the previous one, in that there remains a massless vector field in the final phase; indeed the gauge-field structure is the same as in the bosonic sector of the standard electroweak model.  Since $\pi_1(H/K)=\ZZ$, strings labelled by an integer winding number will be formed in the second transition.  The manifold of vacua becomes $\cM'=G/K=S^3$.  Thus there are no truly stable strings or monopoles in the final phase.  What happens is that each monopole becomes attached to a string; each string is either a closed loop or connects a monopole to an antimonopole.

It is easy to see what a monopole configuration looks like.  At large distance from the monopole, we can take $\vec\ph$ radially outwards, so that
 \beq \ph=\et \fr{x^k}{r}\si^k=\et\bpm\cos\th&\sin\th e^{-i\vp}\\
 \sin\th e^{i\vp}&-\cos\th \epm. \eeq
All around the sphere, $\ps$ must be proportional to the eigenvector with eigenvalue $-1$, so we can take
 \beq \ps=\et'\bpm \sin\fr{\th}{2}e^{-i\vp}\\ -\cos\fr{\th}{2}\epm. \eeq
But it is impossible to make this choice continuous everywhere.  Here it is singular at the south pole, where a string must be attached, around which the phase of $\ps$ changes by $2\pi$.

Note that here the strings carry a magnetic flux $(2\pi/e)$, equal to the magnetic charge on the monopole, so there is only one string attached to each, not two.

{\bf c.}  Very different behaviour can be seen in a model based on the symmetry group $G=SU(3)$ with three fields $\ph,\ps_1,\ps_2$, all in the 8-dimensional adjoint representation \cite{Ng:2008mp}.  In the first stage of symmetry breaking $\ph$ acquires a non-zero expectation value, satisfying $|\ph|=\et$, where $|\ph|^2=\half\tr(\ph^2)$.  The vacuum manifold is then $\cM=SU(3)/U(2)$, which may be identified with the complex projective space $\CC P^2$.  Points in this space may be labelled by triples of complex numbers $Z^T=(z_1,z_2,z_3)$, where $(z_1,z_2,z_3)$ and $(\ka z_1,\ka z_2,\ka z_3)$ represent the same point for any non-zero $\ka\in\CC$.  The point in $\cM$ corresponding to $Z\in\CC P^2$ is
 \beq \ph = \fr{\et}{\sqrt{3}}\left(1-3\fr{ZZ^\dag}{Z^\dag Z}\right). \eeq
For example, we may choose the value
 \beq \ph_0=\et T^8\equiv \fr{\et}{\sqrt{3}}\bpm 1&0&0\\0&1&0\\0&0&-2 \epm, \qquad
 Z_0=\bpm 0\\0\\1 \epm. \label{ph0} \eeq
Here $T_1,\dots,T_8$ are the generators of $SU(3)$, the Gell-Mann matrices \cite{Georgi:1982jb}.

After this first symmetry breaking the remaining symmetry group is $H=U(2) \cong SU(2)\X U(1)/\ZZ_2$.  There are non-trivial loops in $H$, and $\pi_1(H)=\ZZ$, so there are monopoles, labelled by an integer $n$.   But this is somewhat misleading.  Homotopically non-trivial loops in $H$ corresponding to odd values of $n$ cannot lie solely in the $U(1)$ factor; they must include a path in $SU(2)$ from the identity element $\bone$ to $(-\bone)_2 = {\rm diag}(-1,-1,1)$.  A monopole with $n=1$ must in a sense carry a ``$\ZZ_2$ charge'' as well as the monopole charge $2\pi/e$.  Note however that the $\ZZ_2$ charge can only have the values 0 or 1; it obeys the $\ZZ_2$ addition rule, $1+1\equiv0$.  Repeated twice this path in $SU(2)$ is trivial, so for even $n$ the paths can be confined to $U(1)$.  For even $n$, the monopoles do not carry a $\ZZ_2$ charge.

Next, we introduce two more adjoint fields, $\ps_{1,2}$, and choose the potential so that all three fields have 
definite magnitude and are orthogonal, in the sense that 
$\tr(\ph\ps_{1,2})=\tr(\ps_1\ps_2)=0$
and also so that at the minimum $\ps_1$ and $\ps_2$ commute with $\ph$.  
For example, with the choice (\ref{ph0}) for $\ph$, we may take $\ps_{1,2}=\et'T^{1,2}$.  
This then breaks the $SU(2)$ symmetry down to $\ZZ_2$, so the final symmetry group is merely $K=U(1)$.

A typical solution representing the field around a minimal-charge monopole, in spherical polars, is
 \beq \ph=\fr{\et}{2\sqrt{3}}
 \bpm 3\cos\th-1&0&-3\sin\th\, e^{i\vp} \\ 0&2&0 \\ -3\sin\th\, e^{-i\vp}&0&-3\cos\th-1 \epm, \qquad
 Z=\bpm \sin\fr{\th}{2}\,e^{i\vp}\\ 0 \\ \cos\fr{\th}{2} \epm. \label{phmon} \eeq
Suitable forms for the other two fields can be found by starting with $\ps_{1,2}=\et'T^{1,2}$ at the north pole $\th=0$, and applying $SU(3)$ transformations $U(\th,\vp)$ that perform the transformation $\ph(\th,\vp) = U(\th,\ps)\ph_0 U^\dag(\th,\vp)$.  A simple choice is
 \beq U(\th,\vp) = \bpm \cos\fr{\th}{2} & 0 & -\sin\fr{\th}{2}\, e^{-i\vp} \\ 0 & 1 & 0 \\
 \sin\fr{\th}{2}\, e^{i\vp} & 0 & \cos\fr{\th}{2} \epm. \eeq
This means that around the south pole
 \beq U(\pi,\vp) = \bpm 0 & 0 & - e^{-i\vp} \\ 0 & 1 & 0 \\ e^{i\vp} & 0 & 0 \epm. \eeq
Evidently, the configuration of the fields $\ps_{1,2}$ is singular at the south pole.  This singularity cannot be removed by a gauge transformation (though it could of course be moved to a different location), because the path as $\vp$ ranges from 0 to $2\pi$ is non-contractible in $SU(3)$, whereas it would be contractible if $\vp$ ranged from 0 to $4\pi$.  A $\ZZ_2$ string must be attached at the south pole of the monopole configuration.

Every monopole of charge $n=1$ must be attached to a string.  The strings may terminate on monopoles or antimonopoles.  A string may join a pair of equal-charge monopoles or a monopole-antimonopole pair.  However, numerical simulations show that typically the second possibility is much more probable than the first.  If the dynamics leads to the string shortening and disappearing, then in the first case this would lead to charge-2 monopoles, but in the second case to complete annihilation.    The charge-2 monopoles have no $\ZZ_2$ charge, but are pure $U(1)$ monopoles.
 
 {\bf d.} A different choice of potential in the $SU(3)$ model can lead to an alternative symmetry breaking pattern, again with very different behaviour \cite{Heo:1998ms}.  The first stage can proceed as before, with $\ph$ typically given by (\ref{ph0}), breaking the symmetry down to $H=U(2)$ and again generating monopoles.  But then we can choose the potential so that the minimum typically occurs when $\ps_{1,2}=\et'T^{4,6}$, generators that do not commute with $T^8$ and so do not belong to $H$.  This choice has the effect of breaking the symmetry down to $K=\ZZ_3$, the centre of $SU(3)$, comprising the matrices $\{e^{2\pi ni/3}\bone | n=0,1,2\}$.  Consequently, this produces $\ZZ_3$ strings.  Since $K$ is purely discrete, no massless gauge fields remain.
 
Around a typical $\ZZ_3$ string, the fields $\ps_{1,2}$ at large distance behave as
 \beq \ps_1 = \et'(T^4\cos n\vp + T^5\sin n\vp),\qquad \ps_2 = \et'(T^6 \cos n\vp + T^7 \sin n\vp). \eeq
This configuration can be induced by applying the gauge rotation 
 \beq U(\vp) = e^{inT^8\vp/\sqrt{3}} = {\rm diag}(e^{ni\vp/3}, e^{ni\vp/3}, e^{-2ni\vp/3}).
 \label{gaugerotroundstring} \eeq
What then happens around a pre-existing monopole?  As in the previous example, we may expect that at the second symmetry breaking, the two new fields $\ps_{1,2}$ may be frustrated from finding the vacuum configuration everywhere around it.  So we may expect strings to be attached.  But there is an important difference this time.  Here the gauge rotation around a string, Eq.~(\ref{gaugerotroundstring}), does not constitute a closed loop in $SU(3)$ unless $n\equiv 0 \mod 3$, since $U(2\pi) = e^{2\pi ni/3}\bone$.  Consequently, we cannot attach just one $n=1$ string for example to the monopole.  We need three of them, and if we are in the region of parameter space in which forces between identical strings are repulsive, the three will tend to spread out around the monopole.  So this symmetry breaking pattern yields a quite different type of string network, with junctions where three strings meet at a monopole.

Another point should be noted here.  An $n=2$ string may or may not be unstable to splitting into two $n=1$ strings.  But in any case it is topologically equivalent to a $n=-1$ string, {\ie} an $n=1$ string in the opposite direction, so if it is stable to splitting, it is indeed in principle unstable to turning into an $n=-1$ string.  But it may nevertheless be \emph{locally} stable, because this transformation can only happen via the creation of a monpole-antimonopole pair, which requires energy.
 
\section{Monopoles and strings in the standard electroweak model}
\label{stdmodel}
 
 We have already discussed semilocal strings in Sec.~\ref{strings} (see Eq.~(\ref{semilocalstringsoln}))
 in an $SU(2)\times U(1)/\ZZ_2$ model where the $SU(2)$ is global and the $U(1)$ is local. This
 model coincides with the standard model of the electroweak interactions whose symmetry group
is denoted $[SU(2)_L \times U(1)_Y]/ \ZZ_2$, but with the important difference that the $SU(2)_L$ factor 
is gauged. If the $SU(2)_L$ and $U(1)_Y$ coupling constants are denoted by $g$ and $g'$ respectively,
the relative strength of the two coupling constants is given by the ``weak mixing angle'', $\theta_w$, 
defined by
\beq
\tan \theta_w = \frac{g'}{g}
\eeq
with the measured value $\sin^2\theta_w = 0.23$.

Since the semilocal model is the $\sin^2\theta_w \to 1$ limit of the standard model, we expect the 
semilocal string solution to also be present in the standard model. Thus the standard model has an
electroweak string solution given by
 \beq 
 \ph = \et f(\rh) e^{ni\vp}{0\choose 1}, \qquad Z_0=0,\qquad 
 Z_k=-\fr{n}{e\rh} h(\rh) \pa_k\vp.
 \label{electroweakstringsoln}
 \eeq
Note that only the $Z$-gauge field of the standard model is non-vanishing; the charged $W^\pm$
and the electromagnetic gauge fields vanish. Thus this solution is sometimes called a ``$Z$-string''
and also distinguishes it from other embedded electroweak strings called ``$W$-strings'' in which
the $W^\pm$ gauge fields are non-vanishing.

As in the semilocal case, there is no topological reason for the existence of the electroweak string
solution; nor is its existence protected by a topological winding number. Hence we expect the $Z$-string
to be unstable under small perturbations. A detailed stability analysis of the electroweak string shows that 
it is metastable if $m_H < m_Z$ and for $\sin^2\theta_w \gtrsim 0.95$, and is
unstable for other parameters, including the physical values: $m_H =125~{\rm GeV}$, 
$m_Z = 91~{\rm GeV}$, $\sin^2\theta_w = 0.23$.

Since the $Z$-string solution is not topological, a particular $Z$-string can terminate. To understand
the properties of the terminus, we decompose the $Z$-magnetic flux inside the string into a linear
combination of $SU(2)_L$ flux and $U(1)_Y$ flux. When the Higgs has the conventional
vacuum expectation value: $\ph = \eta (0, 1)^T$, the decomposition is
\beq
Z_\mu \equiv \cos\theta_w W_\mu ^3 - \sin\theta_w Y_\mu .
\eeq
The $W^3$ magnetic flux is non-Abelian and can terminate, but the $Y$ magnetic flux is Abelian
and divergenceless, and cannot terminate. Then the $Y$ magnetic flux must extend beyond the
terminus of the $Z$-string and can only do so in the form of massless electromagnetic ($A$) magnetic 
flux defined by
\beq
A_\mu \equiv \sin\theta_w W_\mu^3 + \cos\theta_w Y_\mu .
\eeq
Thus the terminus of the $Z$-string is a source of $A$ magnetic flux {\it i.e.}\ a
magnetic monopole. Note that the $Z$ and $A$ gauge fields are orthogonal, so the magnetic
monopole has ${\rm div}({\bf B}_A) \ne 0$, where ${\bf B}_A$ is the electromagnetic magnetic
field, while it is confined by a string that has nothing to do with electromagnetism. (The situation 
is very similar to the dual case where electrically charged quarks are confined by QCD color strings.)

Before describing the properties of the electroweak magnetic monopole and $Z$-string, we will
provide another way of seeing the existence of the monopole, more in line with the original
paper by Nambu \cite{Nambu:1977ag}. Essentially one constructs a composite adjoint field
\beq
n^a (x) = - \frac{\ph^\dag \tau^a \ph}{\ph^\dag \ph}.
\label{ndefn}
\eeq
Once $\ph$ gets a vacuum expectation value, we will have $n^a \ne 0$. Note that $n^a$ transforms
trivially under $U(1)_Y$ and, as far as its properties under $SU(2)_L$ are concerned, it is exactly
like the field $\ph^a$ in Sec.~\ref{tHPol}. Thus, as in the 't Hooft-Polyakov monopole, we 
can write down a ``hedgehog'' configuration
\beq 
\et n^a = \et f(r)\fr{x^a}{r},\qquad W^a_0=0, \qquad W^a_k = -h(r)\fr{\ep^{akj}x^j}{g r^2}.
\eeq
However, this configuration is disallowed in the underlying model because the relation
in Eq.~(\ref{ndefn}) cannot be inverted to obtain a non-singular $\ph$. Instead there
has to be a string attached to the hedgehog on which $\ph =0$. This is exactly the
location of the $Z$-string.

More explicitly, the asymptotic Higgs and gauge field configurations for an electroweak monopole with a 
semi-infinite $Z$-string along the $-z$ axis are given by
\bea
\ph &=& \et \bpm \cos\theta/2 \cr \sin\theta/2 e^{i\varphi} \epm \label{ewmph} \\
g W_\mu^a &=& - \epsilon^{abc} n^b \partial_\mu n^c + 
         i \cos^2\theta_w n^a (\ph^\dag \partial_\mu \ph - \partial_\mu\ph^\dag \ph) \\
g' Y_\mu &=& -i \sin^2\theta_w (\ph^\dag \partial_\mu \ph - \partial_\mu\ph^\dag \ph).
\eea

A finite segment of $Z$-string will have an electroweak monopole on one end and an
antimonopole on the other end. A field configuration for such a finite energy ``dumbbell'' configuration
can be written as \cite{Nambu:1977ag}
\beq
\ph = \bpm \cos(\Theta/2) \cr \sin(\Theta /2) e^{i\varphi} \epm
\eeq
where 
\beq
\cos\Theta \equiv \cos \theta_m - \cos \theta_{\bar m} + 1
\eeq
and $\theta_m$ and $\theta_{\bar m}$ are the spherical polar angles with the axes origin
located at the monopole and the antimonopole respectively.
Nambu also considered the lifetime of rotating dumbbells, though only accounting for
decay by emission of electromagnetic radiation. In particular, decay by
fragmentation and other instabilities were not considered and remain to be investigated.

The magnetic flux of the electroweak monopole can be shown to be
\beq
F = \frac{4\pi}{e} \sin^2\theta_w.
\eeq
Seemingly this does not obey the Dirac quantization condition but this is not a contradiction
because of the $Z$-string that is attached to the monopole.

The mass of the electroweak monopole cannot be defined because it is always confined.
If the mass is measured in terms of the energy barrier to the breaking of $Z$-strings, it
would turn out to be negative because the $Z$-string is unstable.

Finally we discuss the electroweak ``sphaleron'' \cite{Manton:1983nd} in terms of a bound
state of an electroweak monopole and antimonopole. Since a monopole and an antimonopole
carry opposite magnetic charges, there is an attractive Coulomb force that tends to bring
them together so that they can annihilate. However, a monopole and an antimonopole
have an extra degree of freedom, namely a relative phase between them. To see this in the
context of the electroweak model \cite{Vachaspati:1994ng}, consider the asymptotic Higgs field 
configuration
\beq
\ph_{m{\bar m}} = \et \bpm \sin(\theta_m/2) \sin(\theta_{\bar m}/2) e^{i\gamma} +
                                      \cos(\theta_m/2) \cos(\theta_{\bar m}/2) \cr
                                      \sin(\theta_m/2) \cos(\theta_{\bar m}/2) e^{i\varphi} -
                                      \cos(\theta_m/2) \sin(\theta_{\bar m}/2) e^{i(\varphi - \gamma)} 
                               \epm
\label{phitwisted}
\eeq
where $\theta_m$ and $\theta_{\bar m}$ are spherical polar angles measured from the
location of the monopole at $z =+a$ on the $z-$axis and the location of the antimonopole
at $z=-a$ respectively. The phase angle $\gamma$ will be explained in a moment. Note that
$| \ph_{m{\bar m}}| = \et$.

Away from the antimonopole and close to the monopole we can
take $\theta_{\bar m} \approx 0$ and the configuration reduces to
\beq
\ph_{m{\bar m}} \to \et \bpm  \cos(\theta_m/2) \cr \sin(\theta_m/2) e^{i\varphi}  \epm
\eeq
which is the configuration around an electroweak monopole (see Eq.~(\ref{ewmph})). On
the other hand, if we take $\theta_m \to \pi$,
\beq
\ph_{m{\bar m}} \to \et e^{i\gamma} 
           \bpm  \sin(\theta_{\bar m}/2) \cr \cos(\theta_{\bar m}/2) e^{i(\varphi -\gamma)}  \epm .
\eeq
which is the Higgs configuration around an antimonopole up to an irrelevant overall phase
factor.
The $\varphi-\gamma$ phase shows that the antimonopole has a relative rotation compared to the 
monopole. Thus $\gamma$ is a relative ``twist'' between the monopole and the antimonopole. 
Further, the twist provides a repulsive force between the monopole and the antimonopole. By
adjusting the twist parameter and the monopole-antimonopole separation, a static solution can
be found. The solution was first found in an $O(3)$ model in Ref.~\cite{Taubes:1982ie} and then in the
electroweak model (in the $\theta_w =0$ limit) in Ref.~\cite{Manton:1983nd} using very elegant mathematical 
techniques. The solution, now called a ``sphaleron'', plays an important role in anomalous 
baryon number violation, and may play a critical role in explaining the cosmic matter-antimatter asymmetry,
and may also provide a mechanism to generate cosmological magnetic fields (see below).

 \section{Observational constraints}
 \label{observations} 
 \begin{figure}
  \begin{center}
  \includegraphics[width=120mm]{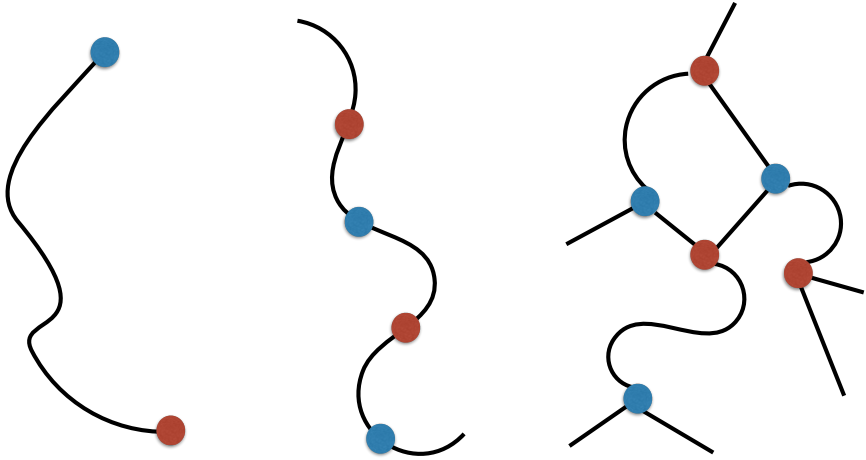}
  \end{center}
\caption{Three cases of monopoles on strings; blue and red dots represent monopoles and antimonopoles respectively.}
\label{3cases}
\end{figure}
 
As depicted in Fig.~\ref{3cases}, there are three distinct cases relevant to monopoles connected by 
strings that need to be considered in a cosmological setting. The three cases correspond to whether a 
monopole is connected by 1 or 2 or many ($\ge 3$) strings. In addition, in all three cases, we can
consider the possibility that {\em{all}} the monopole magnetic flux has been confined to the string, or only
{\em{some}} of the flux is confined while the remaining flux is unconfined. For example, in Sec.~\ref{mono-str}
we have discussed the case of $SU(2) \to U(1) \to 1$ and there all the monopole flux gets confined
to a string. On the other hand, for the electroweak monopole discussed in Sec.~\ref{stdmodel}, the
$Z$-flux is confined to a string, but the monopole still carries an unconfined electromagnetic flux.

First consider the case when a monopole is connected to an antimonopole by a single string and forms
a ``dumbbell''. Simulations find that the length distribution of dumbbells is exponential: $\exp(-l/\xi)$
where $\xi$ is set by the average distance between monopoles at the time of string formation\cite{Copeland:1987ht}.

If dumbbells are produced 
at some cosmological epoch, the strings will quickly shrink and bring the monopole and antimonopole
together. 
The acceleration of the monopole and antimonopole will lead to electromagnetic radiation,
whereby the system will lose energy, as given by the classical electromagnetic radiation formula
$\dot E = g^2 a^2/6\pi$ where the acceleration $a = \mu /M$, $M$ is the monopole mass, 
$g$ the magnetic charge, and $\mu$ the string tension. 
Once the monopole and antimonopole collide, they annihilate and the time scale for a given dumbbell 
to dissipate will be
given by its initial length. If a dumbbell is very long initially, there is also a probability that it will be
chopped into shorter segments when other strings collide or when the long string crosses on itself, 
through a process called intercommutation. Since the length distribution of dumbbells is dominated
by the smallest length, and the smallest length is typically much shorter than the cosmic horizon, most
of the energy in dumbbells is dissipated within a Hubble time. Then observational signatures can 
only arise if a telltale remnant is produced during the decay process. We will shortly discuss 
three possible remnants.

 In the second case shown in Fig.~\ref{3cases}, 
 the monopole is like a ``bead on a string'' and we expect the formation of
 ``cosmic necklaces'' \cite{Hindmarsh:1985xc,Leblond:2007tf}. 
Monopoles and antimonopoles can slide along the string, collide, and annihilate, and produce high energy particles that can potentially
be observed as cosmic rays \cite{Berezinsky:1997td,Siemens:2000ty,BlancoPillado:2007zr}.
However, closer scrutiny of the process \cite{BlancoPillado:2007zr} finds that the monopoles 
annihilate very rapidly after formation and the network soon resembles a network of ordinary 
cosmic strings. Then the observational constraints on ordinary cosmic strings (discussed below) 
also apply to beads-on-strings, independent of whether the monopole (beads) carry unconfined 
magnetic flux.

 In the third case of Fig.~\ref{3cases}, a string web is formed with monopoles at the junctions of the 
 web. Then the web 
 stretches with the expansion of the universe, and dilutes due to monopole-antimonopole annihilation. 
 The resulting network scales self-similarly in time, {\it i.e.} the statistical properties of the web do
 not depend on time but the characteristic overall scale (distance between monopoles) grows in proportion
 to cosmic time \cite{Vachaspati:1986cc}: $d(t) \sim \chi t$ with $\chi \sim \mu/(24\pi M^2)$,
 where $\mu$ is the string tension and $M$ is the monopole mass. We now need to distinguish
 between the case when the monopoles carry unconfined flux and the case when all the flux is confined,
 since these two scenarios lead to very different cosmological scenarios \cite{Vachaspati:1986cc}.
 
  If the monopoles carry unconfined flux, their rapid acceleration under the pull of the strings
 leads to the emission of very high energy gamma rays whose spectrum peaks at  $\sim 100~{\rm TeV}$.
 The energy density in such gamma rays divided by the critical energy density of the universe is
 estimated to be \cite{Vachaspati:1986cc}
 \begin{equation}
 \Omega_{\gamma~{\rm TeV}} \sim \frac{30 G\mu}{\chi^2} \Omega_\gamma
 \end{equation}
 where $\Omega_\gamma$ is the fractional cosmic radiation energy density. The observed gamma ray
 flux dies off very rapidly at such high energies. Using the numerical values in 
 Ref.~\cite{Ackermann:2014usa},
 the relative energy density in cosmic gamma rays at energies above say 100~GeV is
 $\Omega_{\gamma > 100 {\rm GeV}} \lesssim 10^{-11}$, thus leading to the constraint
 \begin{equation}
 \frac{M^2}{m_P \sqrt{\mu}} \lesssim 10^{-6}
 \label{unconfinedwebconstraint}
 \end{equation}
 where $m_P = 1.2\times 10^{19}~{\rm GeV}$ is the Planck mass. The constraint will be stronger
 if we restrict to gamma rays with energy greater than $\sim 100~{\rm TeV}$ where observations indicate 
 a sharp cutoff in the gamma ray flux.
It has been suggested that under some circumstances even particles of trans-Planckian energy could have been generated \cite{Berezinsky:1997kd}.
 
 If all the magnetic flux of the monopoles is confined to the string, the monopoles do not radiate high
 energy photons even as they are accelerated by the connecting strings to relativistic energies. In this
 case, the web of strings and monopoles does not have an efficient way to dissipate its energy. The
 energy density in the web then dilutes due to Hubble expansion and due to occasional rearrangements
 when monopoles annihilate or when strings intercommute. As a result the relative energy density
 in the web compared to the matter density grows with time \cite{Vachaspati:1986cc,Martins:2010ma}. 
 Eventually the web dominates the cosmological matter energy density. Once the cosmological evolution 
 of the web is understood in detail\footnote{And with the inclusion of dark energy.}, the
 growth of the web relative energy density potentially leads to a constraint on the parameters of the 
 fundamental model but a rigorous constraint has not been derived so far.
 
Current cosmological constraints on ordinary cosmic strings as derived from the millisecond pulsar
timing observations limit the mass per unit length of the string, $\mu$, to be less than $\sim 10^{-9}$ 
in Planck units,  {\it i.e.}\ $\mu \lesssim 10^{19}~{\rm gm/cm}$ \cite{Jenet:2006sv,Blanco-Pillado:2013qja}. 
This constraint depends on the gravitational radiation from strings, which is turn depends on the dynamics 
of strings, and in particular on the loop distribution. In the cases of beads on strings, the dynamics is expected 
to be similar to ordinary cosmic strings and so this constraint also applies. However, in the case of a web
of strings, the dynamics is very different and loop formation is suppressed. In this case, the constraints
from the non-detection of string gravitational lensing, and non-observation of string induced
distortions of the angular power spectrum of the cosmic microwave background (CMB), can still
be applied. These provide the bound $\mu \lesssim 10^{-7}$ (for a summary of observational
bounds on cosmic strings, see Ref.~\cite{scholarpediapage}).  If we combine this bound on the string
tension with Eq.~(\ref{unconfinedwebconstraint}) for the case of string webs in which monopoles have 
unconfined gauge flux, we obtain a constraint on the monopole mass
\begin{equation}
M \lesssim 10^{-5} m_P.
\end{equation}
  
 There are two remnants that can arise from dumbbells that are created at some cosmological epoch
that can potentially lead to an observable signature. 
 The first remnant is simply the energy resulting from the decay of dumbbells provided they decay
 at cosmological redshifts between $z \approx 10^4$ and $10^6$. In this case there is not enough time 
 left until hydrogen recombination for the energy injected into the cosmological medium to get thermalized. 
 As a result, the decay of dumbbells can distort the spectrum of the CMB. No such distortions have been 
 measured so far and this limits the amount of energy deposition in the medium. 
 However, the cosmic temperature at these redshifts is $< 1~{\rm keV}$ and the cosmic
 time is $\sim 1~{\rm yr}$, and from the particle physics side, we think we know that there are no dumbbells
 that can survive for this long a period. The exception is if plasma effects can somehow play a role as
 discussed in the case of ``embedded defects'' in Ref.~\cite{Nagasawa:2002at,Karouby:2012yz}
 or if quantum effects are important and stabilize the dumbbells \cite{Weigel:2015lva}.
  
 The second remnant produced by decaying dumbbells is a magnetic field that can be trapped in
 the cosmological medium, which can then survive until the present 
 epoch \cite{Vachaspati:1991nm,Vachaspati:1994xc}.  Indeed, primordial magnetic fields 
 may also help explain the ubiquity of magnetic fields seen in galaxies and clusters of galaxies
 (for a recent review, see \cite{Subramanian:2015lua}). 

 We have already related twisted dumbbells to the electroweak sphaleron in Sec.~\ref{stdmodel}
 (see around Eq.~(\ref{phitwisted})). If we assume that the cosmic matter-antimatter asymmetry is 
 generated dynamically via sphaleron processes, then sphaleron decay will leave behind twisted
 or ``helical'' magnetic fields \cite{Copi:2008he,Chu:2011tx}.
 Such magnetic fields violate parity since the handedness of the field is
 related to the preference of matter over antimatter. Evidence for helical cosmological 
 magnetic fields has recently been discovered \cite{Tashiro:2013ita, Chen:2014qva}, suggesting that 
 they may have been produced during the decay of monopoles-on-strings. 
 
 If the monopoles on dumbbells do not carry unconfined flux, they will still lose energy by emitting 
 gravitational waves, thus providing a third cosmological remnant from dumbbells. Further, if the 
 dumbbells are sufficiently long at production, as can happen if the strings are produced
 after an inflationary epoch or with certain string theory cosmic strings, the distribution of
 dumbbells will produce a gravitational wave background \cite{Martin:1996ea} and
 gravitational wave bursts \cite{Leblond:2009fq}. Upcoming gravitational wave detectors can be
 sensitive to the bursts and can potentially provide constraints at the level $G\mu \lesssim 10^{-12}$.
  
 \section{Conclusions}\label{conclusions}

Field theories admit a wide variety of topological defects of which monopoles, strings, and domain walls
are commonly discussed. In this review we have focussed on a type of ``hybrid'' or ``composite'' defect,
namely monopoles connected by strings. We have discussed field theories in which monopoles are 
connected to 1, 2, or 3 strings. The case of one string per monopole is relevant to the electroweak model,
and also to a proposed explanation of the observed absence of cosmological magnetic monopoles
\cite{Langacker:1980kd}. For more than 1 string per monopole, we have considered the symmetry breaking
pattern $SU(N) \to SU(N-1)\times U(1) \to \ZZ_N$. The first stage of symmetry breaking gives monopoles 
and the second connects the monopoles to $N$ strings. We have also described the monopoles
connected by a single string arising in the symmetry breaking $SU(2)\times U(1) \to U(1)\times U(1)
\to U(1)$ and this is directly relevant to the standard electroweak model.

Monopoles-on-strings can have observable effects in cosmology and ongoing observational
efforts constrain their abundance. If monopoles are connected by 2 or more strings, a string
network should exist in the universe. The strongest bounds on a string network arise from
gravitational radiation from loops of strings and lead to $\mu \lesssim 10^{-9}$, where
$\mu$ is the string tension in Planck units. The bound may not apply to the string web in which 
monopoles are connected by more than 2 strings, since the loop distribution will likely be
suppressed. Gravitational lensing constraints still imply $\mu \lesssim 10^{-7}$. Non-gravitational
constraints due to particle emission have also been derived in the literature
and are summarized  in Sec.~\ref{observations}.

The case when a monopole is connected by a single string is special because the strings then
bring monopoles and antimonopoles together, and the whole system can rapidly annihilate.
In this case, cosmological observables can only be sensitive to the decay products of the
system. Since the annihilation of monopoles and antimonopoles releases magnetic fields,
the growing evidence for cosmological magnetic fields may indeed indicate a role for
monopoles-on-strings in the early universe.

\section*{Acknowledgements}

This work was supported by the Department of Energy at Arizona State University.


\section*{References}

\begin{thebibliography}{10}

\bibitem{Amaldi:1991zx}
U.~Amaldi, W.~de Boer, P.~H.~Frampton, H.~Furstenau and J.~T.~Liu,
Phys.\ Lett.\ B {\bf 281}, 374 (1992).

\bibitem{Haber:1997if}
H.~E.~Haber,
Nucl.\ Phys.\ Proc.\ Suppl.\  {\bf 62}, 469 (1998),
[arXiv:hep-ph/9709450].

\bibitem{Mambrini:2015vna} 
  Y.~Mambrini, N.~Nagata, K.~A.~Olive, J.~Quevillon and J.~Zheng,
  Phys.\ Rev.\ D {\bf 91}, no. 9, 095010 (2015)
  [arXiv:1502.06929 [hep-ph]].

\bibitem{Sarangi:2002yt} 
  S.~Sarangi and S.~H.~H.~Tye,
  Phys.\ Lett.\ B {\bf 536}, 185 (2002)
  [hep-th/0204074].

\bibitem{Jeannerot:2003qv}
  R.~Jeannerot, J.~Rocher and M.~Sakellariadou,
  Phys.\ Rev.\ D {\bf 68} (2003) 103514
  [hep-ph/0308134].

\bibitem{Jones:2002cv}
  N.~T.~Jones, H.~Stoica and S.~H.~H.~Tye,
  JHEP {\bf 0207} (2002) 051
  [hep-th/0203163].

\bibitem{Majumdar:2002hy}
  M.~Majumdar and A.-C.~Davis,
  JHEP {\bf 0203} (2002) 056
  [hep-th/0202148].

\bibitem{Becker:2005pv}
  K.~Becker, M.~Becker and A.~Krause,
  Phys.\ Rev.\ D {\bf 74} (2006) 045023
  [hep-th/0510066].

\bibitem{Jackson:2004zg}
  M.~G.~Jackson, N.~T.~Jones and J.~Polchinski,
  JHEP {\bf 0510} (2005) 013
  [hep-th/0405229].
  
\bibitem{Hashimoto:2005hi}
  K.~Hashimoto and D.~Tong,
  JCAP {\bf 0509} (2005) 004
  [hep-th/0506022].

\bibitem{Polchinski:2004ia}
  J.~Polchinski,
  Lectures presented at the 2004 Carg\`ese Summer School
  [hep-th/0412244].
 
\bibitem{Dvali:2003zj}
  G.~Dvali and A.~Vilenkin,
  JCAP {\bf 0403} (2004) 010
  [hep-th/0312007].

\bibitem{Avgoustidis:2005nv}
  A.~Avgoustidis and E.~P.~S.~Shellard,
  Phys.\ Rev.\ D {\bf 73} (2006) 041301
  [astro-ph/0512582].
 
\bibitem{Tye:2005fn}
  S.-H.~H.~Tye, I.~Wasserman and M.~Wyman,
  Phys.\ Rev.\ D {\bf 71} (2005) 103508
   [Phys.\ Rev.\ D {\bf 71} (2005) 129906]
  [astro-ph/0503506].
 
\bibitem{Hindmarsh:2006qn}
  M.~Hindmarsh and P.~M.~Saffin,
  JHEP {\bf 0608} (2006) 066
  [hep-th/0605014].

\bibitem{Kneipp:2007fg} 
  M.~A.~C.~Kneipp,
  Phys.\ Rev.\ D {\bf 76}, 125010 (2007)
  [arXiv:0707.3791 [hep-th]].

\bibitem{Sakai:2005sp}
  N.~Sakai and D.~Tong,
  JHEP {\bf 0503} (2005) 019
  [hep-th/0501207].

\bibitem{Tong:2005un}
  D.~Tong,
  TASI lectures on solitons: {\it Instantons, monopoles, vortices and kinks}
  [hep-th/0509216].

\bibitem{Kibble:2002ex}
  T.~W.~B.~Kibble,
  in \emph{Patterns of Symmetry Breaking}, ed. H. Arodz, J. Dziarmaga \& W.H. Zurek, NATO 
  Science II,  {\bf 127}, pp. 3--36 (2002).                    
  [arXiv:cond-mat/0211110].

\bibitem{Turok:1989ai} 
  N.~Turok,
  Phys.\ Rev.\ Lett.\  {\bf 63}, 2625 (1989).

\bibitem{Nielsen:1973cs}
  H.~B.~Nielsen and P.~Olesen,
  Nucl.\ Phys.\ B {\bf 61} (1973) 45.

\bibitem{Achucarro:1999it}
A.~Achucarro and T.~Vachaspati,
\newblock Phys.\ Rept.\ {\bf 327}, 347 (2000), arXiv:hep-ph/9904229.

\bibitem{Hindmarsh:1991jq} 
  M.~Hindmarsh,
  Phys.\ Rev.\ Lett.\  {\bf 68}, 1263 (1992).

\bibitem{James:1992wb} 
  M.~James, L.~Perivolaropoulos and T.~Vachaspati,
  Nucl.\ Phys.\ B {\bf 395}, 534 (1993)
  [hep-ph/9212301].

\bibitem{'tHooft:1974qc}
  G.~'t Hooft,
  Nucl.\ Phys.\ B {\bf 79} (1974) 276.
 
\bibitem{Polyakov:1974ek}
  A.~M.~Polyakov,
  JETP Lett.\  {\bf 20} (1974) 194
   [Pisma Zh.\ Eksp.\ Teor.\ Fiz.\  {\bf 20} (1974) 430].

\bibitem{Bogomolnyi:1976}
 E.~B.~Bogomol'nyi,
 Sov.\ J.\ Nucl.\ Phys.\ {\bf 24}, 449 (1976).

\bibitem{Prasad:1975kr}
  M.~K.~Prasad and C.~M.~Sommerfield,
  Phys.\ Rev.\ Lett.\  {\bf 35} (1975) 760.
 
\bibitem{Kibble:1982dd}
  T.~W.~B.~Kibble, G.~Lazarides and Q.~Shafi,
  Phys.\ Rev.\ D {\bf 26} (1982) 435.

\bibitem{Leblond:2007tf}
  L.~Leblond and M.~Wyman,
  Phys.\ Rev.\ D {\bf 75} (2007) 123522
  [astro-ph/0701427].

\bibitem{Ng:2008mp}
  Y.~Ng, T.~W.~B.~Kibble and T.~Vachaspati,
  Phys.\ Rev.\ D {\bf 78} (2008) 046001
  [arXiv:0806.0155 [hep-th]].

\bibitem{Georgi:1982jb}
  H.~Georgi,
  Front.\ Phys.\  {\bf 54} (1982) 1.

\bibitem{Heo:1998ms}
  J.~Heo and T.~Vachaspati,
  Phys.\ Rev.\ D {\bf 58} (1998) 065011
  [hep-ph/9801455].

\bibitem{Nambu:1977ag} 
  Y.~Nambu,
  Nucl.\ Phys.\ B {\bf 130}, 505 (1977).
  
\bibitem{Manton:1983nd} 
  N.~S.~Manton,
  Phys.\ Rev.\ D {\bf 28}, 2019 (1983).
  
\bibitem{Vachaspati:1994ng} 
  T.~Vachaspati and G.~B.~Field,
  Phys.\ Rev.\ Lett.\  {\bf 73}, 373 (1994)
  [hep-ph/9401220].
  
\bibitem{Taubes:1982ie} 
  C.~H.~Taubes,
  Commun.\ Math.\ Phys.\  {\bf 86}, 257 (1982).
 
\bibitem{Copeland:1987ht} 
  E.~J.~Copeland, D.~Haws, T.~W.~B.~Kibble, D.~Mitchell and N.~Turok,
  Nucl.\ Phys.\ B {\bf 298}, 445 (1988).

 \bibitem{Hindmarsh:1985xc} 
  M.~Hindmarsh and T.~W.~B.~Kibble,
  Phys.\ Rev.\ Lett.\  {\bf 55}, 2398 (1985).

\bibitem{Berezinsky:1997td} 
  V.~Berezinsky and A.~Vilenkin,
  Phys.\ Rev.\ Lett.\  {\bf 79}, 5202 (1997)
  [astro-ph/9704257].

\bibitem{Siemens:2000ty} 
  X.~Siemens, X.~Martin and K.~D.~Olum,
  Nucl.\ Phys.\ B {\bf 595}, 402 (2001)
  [astro-ph/0005411].

\bibitem{BlancoPillado:2007zr} 
  J.~J.~Blanco-Pillado and K.~D.~Olum,
  JCAP {\bf 1005}, 014 (2010)
  [arXiv:0707.3460 [astro-ph]].

\bibitem{Vachaspati:1986cc} 
  T.~Vachaspati and A.~Vilenkin,
  Phys.\ Rev.\ D {\bf 35}, 1131 (1987).

\bibitem{Ackermann:2014usa} 
  M.~Ackermann {\it et al.}  [The Fermi LAT Collaboration],
  Astrophys.\ J.\  {\bf 799}, no. 1, 86 (2015)
  [arXiv:1410.3696 [astro-ph.HE]].

\bibitem{Berezinsky:1997kd} 
  V.~Berezinsky, X.~Martin and A.~Vilenkin,
  Phys.\ Rev.\ D {\bf 56}, 2024 (1997)
  [astro-ph/9703077].

\bibitem{Martins:2010ma} 
  C.~J.~A.~P.~Martins,
  Phys.\ Rev.\ D {\bf 82}, 067301 (2010)
  [arXiv:1009.1707 [hep-ph]].

\bibitem{Jenet:2006sv} 
  F.~A.~Jenet, G.~B.~Hobbs, W.~van Straten, R.~N.~Manchester, M.~Bailes, J.~P.~W.~Verbiest, R.~T.~Edwards and A.~W.~Hotan {\it et al.},
  Astrophys.\ J.\  {\bf 653}, 1571 (2006)
  [astro-ph/0609013].
  
\bibitem{Blanco-Pillado:2013qja} 
  J.~J.~Blanco-Pillado, K.~D.~Olum and B.~Shlaer,
  Phys.\ Rev.\ D {\bf 89}, no. 2, 023512 (2014)
  [arXiv:1309.6637 [astro-ph.CO]].
 
 \bibitem{scholarpediapage}
 Tanmay Vachaspati, Levon Pogosian, and Dani\`ele~A. Steer, (2015) {\it Cosmic strings}. Scholarpedia, 10(2):31682.
  \url{http://www.scholarpedia.org/article/Cosmic_strings}

\bibitem{Nagasawa:2002at} 
  M.~Nagasawa and R.~Brandenberger,
  Phys.\ Rev.\ D {\bf 67}, 043504 (2003)
  [hep-ph/0207246].

\bibitem{Karouby:2012yz} 
  J.~Karouby and R.~Brandenberger,
  Phys.\ Rev.\ D {\bf 85}, 107702 (2012)
  [arXiv:1203.0073 [hep-th]].

\bibitem{Weigel:2015lva} 
  H.~Weigel, M.~Quandt and N.~Graham,
  [arXiv:1505.07631 [hep-th]].

\bibitem{Vachaspati:1991nm} 
  T.~Vachaspati,
  Phys.\ Lett.\ B {\bf 265}, 258 (1991).

\bibitem{Vachaspati:1994xc} 
  T.~Vachaspati,
  Sintra Electroweak 1994:171-184
  [hep-ph/9405286].

\bibitem{Subramanian:2015lua} 
  K.~Subramanian,
  {\it The origin, evolution and signatures of primordial magnetic fields},
  arXiv:1504.02311 [astro-ph.CO].

\bibitem{Copi:2008he} 
  C.~J.~Copi, F.~Ferrer, T.~Vachaspati and A.~Achucarro,
  Phys.\ Rev.\ Lett.\  {\bf 101}, 171302 (2008)
  [arXiv:0801.3653 [astro-ph]].

\bibitem{Chu:2011tx} 
  Y.~Z.~Chu, J.~B.~Dent and T.~Vachaspati,
  Phys.\ Rev.\ D {\bf 83}, 123530 (2011)
  [arXiv:1105.3744 [hep-th]].

\bibitem{Tashiro:2013ita} 
  H.~Tashiro, W.~Chen, F.~Ferrer and T.~Vachaspati,
  Monthly Notices of the Royal Astronomical Society: Letters 2014
  445 (1): L41-L45
  [arXiv:1310.4826 [astro-ph.CO]].
  
\bibitem{Chen:2014qva} 
  W.~Chen, B.~D.~Chowdhury, F.~Ferrer, H.~Tashiro and T.~Vachaspati,
  {\it Intergalactic magnetic field spectra from diffuse gamma rays},
  arXiv:1412.3171 [astro-ph.CO].

\bibitem{Langacker:1980kd} 
  P.~Langacker and S.~Y.~Pi,
  Phys.\ Rev.\ Lett.\  {\bf 45}, 1 (1980).

\bibitem{Martin:1996ea} 
  X.~Martin and A.~Vilenkin,
  Phys.\ Rev.\ Lett.\  {\bf 77}, 2879 (1996)
  [astro-ph/9606022].

\bibitem{Leblond:2009fq} 
  L.~Leblond, B.~Shlaer and X.~Siemens,
  Phys.\ Rev.\ D {\bf 79}, 123519 (2009)
  [arXiv:0903.4686 [astro-ph.CO]].


\end{thebibliography}
\end{document}